\documentclass[pra,reprint, twocolumn, subscriptaddress, showpacs]{revtex4}
\usepackage{amsmath}
\usepackage{graphicx}
\usepackage{hyperref}
\begin{document}

%\preprint{}

\title{Radio frequency spectroscopy measurement of the Land\'e  $g$ factor of the $5D_{5/2}$ state of Ba$^+$ with a single trapped ion}



\author{Matthew R. Hoffman}
\email[]{Corresponding author: hoffman2@u.washington.edu}
\thanks{Authors contributed equally}
%\thanks{}
\author{Thomas W. Noel}
\thanks{Authors contributed equally}
\author{Carolyn Auchter}
\author{Anupriya~Jayakumar}
\author{Spencer R. Williams}
\author{Boris B. Blinov}
\author{E. N. Fortson}
\affiliation{Department of Physics, University of Washington, Seattle, Washington 98195, USA}


\date{\today}

\begin{abstract}
We report an improved measurement of the Land\'e  $g$ factor of the $5D_{5/2}$ state of singly ionized barium.  Measurements were performed on single Doppler-cooled $^{138}$Ba$^+$ ions in  linear Paul traps using two similar, independent apparatuses.  Transitions between Zeeman sublevels of the  $6S_{1/2}$ and $5D_{5/2}$ states were driven with two independent, stabilized radio-frequency synthesizers using a dedicated electrode within each ion trap chamber.  State detection within each Zeeman manifold was achieved with a frequency-stabilized fiber laser operating at 1.76 $\mu$m.  By calculating the ratio of the two Zeeman splittings, and using the measured Land\'e  $g$ factor of the $6S_{1/2}$ state, we find a value of 1.200~371$(4_{stat})(6_{sys})$ for $g_{D_{5/2}}$.
\end{abstract}


%\pacs{32.60.+i,32.10.Fn,32.30.Bv}
%\keywords{}

\maketitle
\section{Introduction}

As atomic theorists' computational techniques become increasingly accurate and sophisticated, precision experiments are necessary to confirm the results of their calculations.  The substructure of the long-lived $5D_{5/2}$ state of Ba$^{+}$ ($\tau \approx 32$s \cite{dehmelt86, gurrell07}) is an ideal testing ground since calculations are complicated by uncertainty in the wavefunctions used.   For this state, the Land\'e $g$ factor is expected to be 6/5 from pure L-S coupling, however QED and relativistic corrections to this quantity have not been predicted by atomic theorists to our knowledge.  A method similar to Ref.~\cite{lindroth93} could offer an improved prediction.  A precision measurement of this quantity could offer insight into these higher order corrections.  

Additionally, relativistic coupled cluster (RCC) calculations can be employed to calculate hyperfine structure constants \cite{sahoo06quad}, and an experiment \cite{lewty12} has measured hyperfine constants in the $5D_{3/2}$ state of $^{137}$Ba$^+$ ($I=3/2$) to the highest precision to date.  If the hyperfine constants are measured in the $5D_{5/2}$ level as well, the nuclear magnetic octupole moment of $^{137}$Ba$^+$ can be extracted unambiguously, as all second-order theory corrections can be eliminated \cite{howell08}.  However, the situation for the $5D_{5/2}$ level is complicated by the fact that the strength of the hyperfine interaction is on the same order as that of the Zeeman interaction in a magnetic field of convenient size.  Thus, to measure the hyperfine constants with sufficient accuracy to test atomic theory, the Land\'e  $g$ factor of the $5D_{5/2}$ level of Ba$^+$, $g_{D_{5/2}}$,  must be measured to higher accuracy than previous measurements \cite{moore1908, back1923, curry04, kurz10}.  

\section{Experimental Procedure}
For this measurement of $g_{D_{5/2}}$, we performed precision radio-frequency (rf) spectroscopy on single trapped $^{138}$Ba$^+$ ions using two independent experimental apparatuses.   The ions are  confined with linear Paul traps, one similar to \cite{olmschenk07}, and the other to \cite{gulde03}, driven with an rf potential operating at approximately 11.4 MHz and 11.9 MHz respectively.  We will refer to these as the ``rod'' trap, and the ``blade" trap in the following discussion, owing to their designs.  A pair of current-carrying coils generates a stable, adjustable magnetic field of up to 10 Gauss, which provides a quantization axis for the ion, as well as the Zeeman splitting of the levels.

\begin{figure}[h!!]
\includegraphics[width=2.6in]{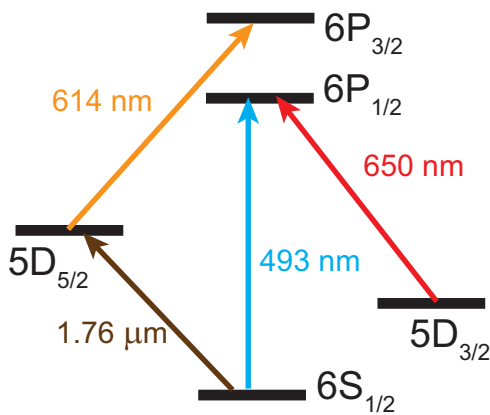}
\caption{Energy levels and relevant transitions in $^{138}$Ba$^+$. The ion is laser cooled on the 493 nm transition and is repumped from the long lived $5D_{3/2}$ state with 650 nm light.
The ion can be shelved using 1.76 $\mu$m light to the metastable $5D_{5/2}$~($m=-5/2$) Zeeman sublevel, where rf spectroscopy can be performed.  A pulse of 614 nm light returns the ion to the cooling cycle via the $6P_{3/2}$ state.}
\label{fig:ba_138_transitions}
\end{figure}

The energy level diagram along with the relevant transitions for $^{138}$Ba$^+$ is shown in Fig.~\ref{fig:ba_138_transitions}. The ion is Doppler cooled on the $6S_{1/2} \leftrightarrow 6P_{1/2}$ transition with a laser operating at 493 nm.  Since the $6P_{1/2}$ state can decay into the long lived $5D_{3/2}$  state ($\tau \approx$ 80s \cite{gurrell07}), the ion must be repumped using a second laser at 650 nm.  Both of these beams are linearly polarized for Doppler cooling.  The ion can be optically pumped into either $6S_{1/2}$ Zeeman sublevel with $>95\%$ efficiency by switching from a cooling 493 nm beam to a second, circularly-polarized beam which is aligned parallel to the quantization axis.

A narrowband fiber laser operating at 1.76 $\mu$m is frequency stabilized to a reference cavity with a finesse of approximately 1000 as detailed in~\cite{kurz10}.  This laser is used to drive the transition from the $6S_{1/2}$~($m_J=-1/2$) ground state sublevel to the $5D_{5/2}$~($m_J=-5/2$) metastable sublevel using adiabatic rapid passage sweeps \cite{noel12,wunderlich07}.  Exciting the ion to the $5D_{5/2}$  shelved  state removes the ion from the cooling cycle \cite{dehmelt86}.  State detection is performed using a photomultiplier tube (PMT) to count photons emitted from the $6P_{1/2} \rightarrow 6S_{1/2}$ transition while the cooling lasers are incident on the ion: a shelved ion will appear ``dark," unshelved will be ``bright."  The ion is returned to the cooling cycle using 614 nm light that addresses the $5D_{5/2} \leftrightarrow 6P_{3/2}$ transition in the ``rod'' trap, while in the ``blade" trap,  the 1.76 $\mu$m laser is used.

Within each ion trap chamber, a dedicated current loop is used to generate tunable rf magnetic fields, which can drive magnetic dipole (M1) transitions between Zeeman sublevels.  An rf synthesizer generates a stable sinusoidal voltage which is amplified and dropped over a 50 ohm rf resistor, producing a current that is passed through the current loop to ground.  The synthesizer in the ``rod" setup was referenced to a SymmetriCom Cs frequency standard;  the synthesizer in the ``blade" setup demonstrated agreement to the Hz level when compared to the same frequency standard.

The experimental sequence for the measurement of the $5D_{5/2}$ Zeeman splitting is as follows.  We first Doppler cool the ion on the 493 nm and 650 nm transitions for approximately 20 ms.  The main 493 nm light is extinguished, and the secondary 493 nm beam with circular polarization $\sigma^-$ optically pumps the ion into the $6S_{1/2}$~($m=-1/2$) sublevel in 10 $\mu$s .  The 493 nm and 650 nm beams are then fully extinguished, and the ion is driven to the $5D_{5/2}$~($m=-5/2$) sublevel using an adiabatic rapid passage sweep with the 1.76 $\mu$m laser.  Typical efficiency for the adiabatic rapid passage is $>90$\% in the ``blade" trap and $\approx$ 80\% in the ``rod" trap.   A constant frequency pulse of rf magnetic field, triggered on a specific phase of the 60 Hz ac voltage is then applied.  Then the 1.76 $\mu$m laser is swept again.  If a transition between the $5D_{5/2}$ Zeeman sublevels was driven by the rf pulse, the ion would remain in the $5D_{5/2}$ state and would be ``dark" when the cooling lasers are turned on.  If no transition occurred, the ion would be returned to the $6S_{1/2}$ state and appear ``bright."  The fluorescence state of the ion is recorded, and, if the ion was ``dark," the ion is returned to the cooling cycle.    The experiment is then repeated, varying the frequency of the rf magnetic field, until a resonance frequency, $f_{D_{5/2}}$, is found.  In a similar way, the Zeeman transition frequency of the ground state, $f_{S_{1/2}}$, is measured.  

To calculate the Land\'e  $g$ factor of the $5D_{5/2}$ state, $f_{S_{1/2}}$ and $f_{D_{5/2}}$ must be measured in quick succession to counter the effects of fluctuations in the ambient magnetic field.  The resonance frequency of each state is given by:
\begin{equation}
f_{i} = \frac{1}{h} g_{i} \mu_B B
\end{equation}
where $h$ is Planck's constant, $\mu_B$ is the Bohr magneton, and $B$ is the magnitude of the laboratory magnetic field. Given measured resonance frequencies, $g_{D_{5/2}}$ can be calculated from:
\begin{equation}
g_{D_{5/2}} = g_{S_{1/2}} \frac{f_{D_{5/2}}}{f_{S_{1/2}}},
\end{equation}
assuming that the magnetic field remains constant over the duration of the experiment.  The experimentally measured value of $g_{S_{1/2}}$ is 2.002 490 6(11)  \cite{knab93}.

\begin{figure}[h!!]
\includegraphics[width=3.4in]{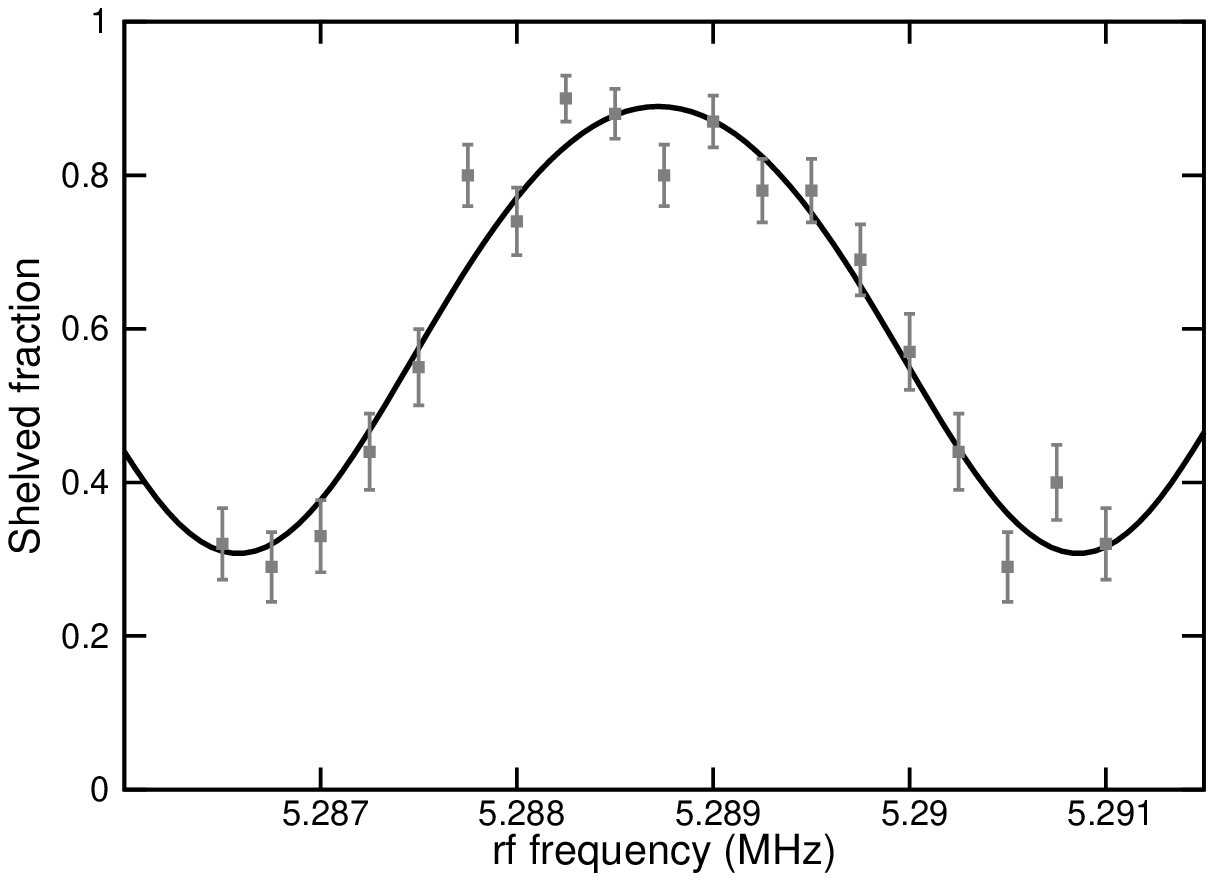}
\includegraphics[width=3.4in]{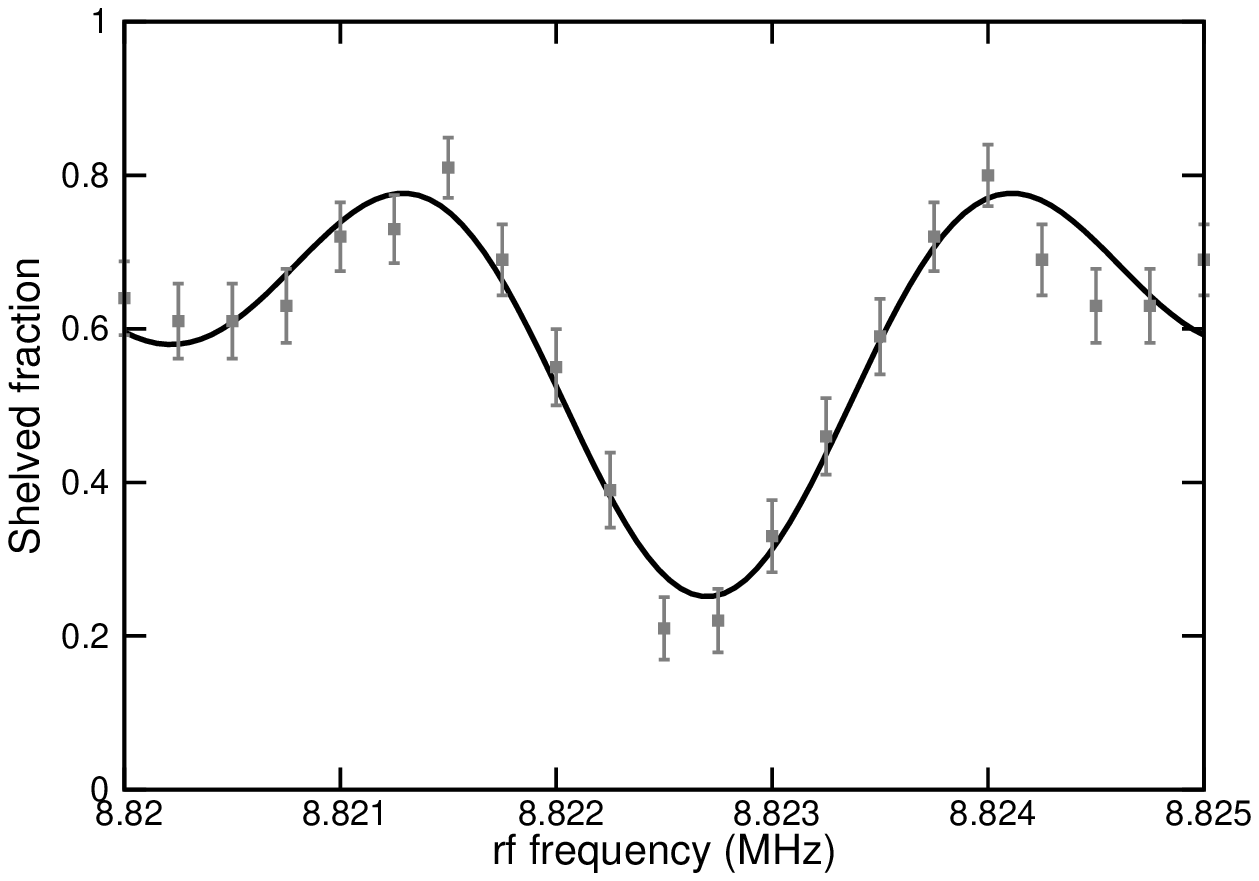}
\caption{Typical experimental data showing the shelved fraction of the ions plotted against the frequency of the oscillating magnetic field.  The upper plot shows the $5D_{5/2}$ resonance, and the bottom shows the $6S_{1/2}$ resonance.  Statistical error bars are shown, calculated from a binomial distribution based on 100 trials at each frequency.  A least-squares fit function based on Eq.~\eqref{eq:fit} is overlaid.}
\label{fig:shelving}
\end{figure}

We measured the resonance frequencies, $f_{S_{1/2}}$ and $f_{D_{5/2}}$, at an applied field of approximately 3-4 Gauss in each setup.  A typical measurement of the ground and exited states ``shelving" probability is shown in Fig.~\ref{fig:shelving}.  The fraction of the trials finding an ion in the ``dark" state is plotted against the frequency of the applied rf current, with error bars indicating the 1-$\sigma$ statistical standard deviations based on 100 trials at each frequency.  The curves are found from four parameter least squares fits to the function:

\begin{equation} \label{eq:fit}
P(f)=\alpha + \beta \frac{\Omega^2}{W^2} \sin^2\left[ W \frac{t}{2}     \right] 
\end{equation}
where $W^2 = \Omega^2 + \left(2 \pi (f-f_0)\right)^2$.  The four fit parameters are two scaling factors, $\alpha$ and $\beta$, which account for imperfect transfer efficiency with the 1.76 $\mu$m laser and imperfect optical pumping, the Rabi frequency of the transition, $\Omega$, and the resonance frequency, $f_0$.

\begin{figure}[h!!]
\includegraphics[width=3.4in]{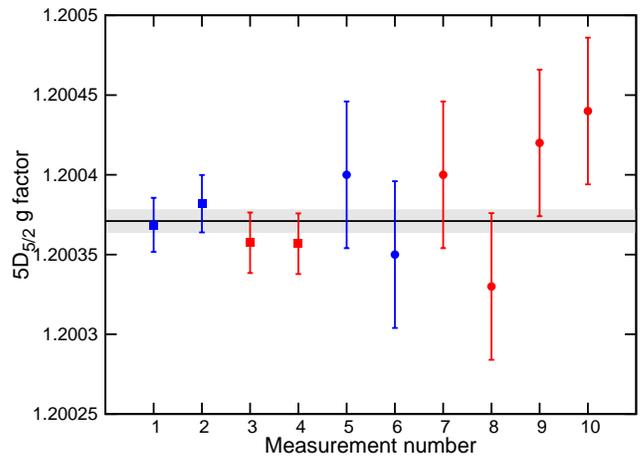}
\caption{Summary of measurements of the Land\'e  $g$ factor of $5D_{5/2}$.  The 10 measurements of $g_{D5/2}$ are shown with squares for the ``rod" trap apparatus, and circles for the ``blade" trap setup.  The blue and red colors indicate different trap depths, which should manifest an ac-Zeeman effect. The error bars represent calculated 1-$\sigma$ total uncertainties (statistical and systematic).  As discussed in the text, the error bars of measurements made in the ``blade" trap are larger as a result of the larger measured magnetic field fluctuations in that apparatus.  Even so, the mean values measured in each apparatus agree within 1-$\sigma$. A solid line represents the weighted mean of all of the measurements done in both apparatuses, with a 1-$\sigma$ confidence interval.}
\label{fig:gmeas}
\end{figure}

Each pair of independent resonance measurements allows us to calculate $g_{D_{5/2}}$.  The results of these measurements are summarized in Fig.~\ref{fig:gmeas}, shown with a combined statistical uncertainty and systematic uncertainty. The statistical uncertainty arises from the binomial uncertainties in the measured shelved fractions at each frequency propagated through to an error in the fit resonance frequency, $f_0$, and the systematic uncertainty will be discussed in the following section.

\section{Systematic Effects}

There are several systematic effects that could skew the measurement of $g_{D_{5/2}}$, however all appear to be well controlled in our experiment.  The effects that we analyze are magnetic field fluctuation, magnetic field gradients, ac Zeeman effect, and 60 Hz jitter.

%The first, due to the presence of 60 Hz ac magnetic fields from equipment in our laboratory could have two effects, namely that it will broaden each resonance due to the 500 $\mu$s pulse duration, and that there could be timing error on the line trigger.  Using the fluxgate magnetometer, as well as the ion's ground state resonance, we find that the magnetic field changes by approximately 5 mG peak to peak over the ac phase.  One experiment is triggered on the zero crossing, resulting in a 150 Hz broadening the ground state resonance, however this is insignificant compared to the $\approx$1 kHz full width at half maximum shown in Fig.~\ref{fig:shelving}.  This will not skew the $g$ factor measurement, though.  There is, however, some slight timing error, that was measured to be approximately 20 $\mu$s, which would shift the ground state resonance frequencies by at most 10 Hz.  However, this timing jitter would have to be systematically off for the excited state Zeeman splitting measurement, and we observe no evidence of this occurring.  Thus, we can safely assume that this fractional error is below the $10^{-7}$ level.

Because neither experimental apparatus employs magnetic shielding, fluctuating ambient magnetic field is the most important systematic effect.  We are aware of several sources for this error.  The largest is due to the public transit system in close proximity to the experiments.  Large unbalanced currents powering electric buses transiently change the magnetic field in the laboratory.  Additionally, we have observed magnetic field drifts as magnetized objects in the laboratory slowly relax.

 To quantify this effect, we proceeded in two directions.  Since the ground state splitting, $f_{S_{1/2}}$, is proportional to the magnetic field, any change in the ambient field would change the resonant splitting.  By repeating the measurement of $f_{S_{1/2}}$, we can observe any fluctuations in the magnetic field.  Doing so, we find that the magnetic field fluctuates randomly by approximately 180  $\mu$G over a 5 minute period for the ``blade" setup.  This appears as an uncertainty in $g_{D_{5/2}}$ of $4.6\times10^{-5}$. Additionally, we installed a three-axis fluxgate magnetometer near the ``rod" trap and monitored the magnetic field while data was being collected.  We find that the largest drift over the course of a measurement of the Zeeman resonance (approximately 400 s) corresponds to a 20 $\mu$G drift in the ambient field resulting in a conservative systematic error of $1.3\times10^{-5}$ to  each measurement of $g_{D_{5/2}}$.  Since the magnetic field fluctuates randomly, multiple measurements will reduce this error.  Any fast fluctuations of the magnetic field will result in broadening of the Zeeman resonance line shape.

A magnetic field gradient could also affect this measurement if the ion were to shift its position over the course of the experiment.  To check this, we measured the ground state resonance, $f_{S_{1/2}}$, at two positions along the trap axis separated by 26 $\mu$m.  The observed change in the resonance frequency corresponded to a change in the magnetic field of 360 $\mu$G, indicating a 0.14 G/cm gradient.  However, we find that during a single experiment, the ion does not change its position by more than 1 $\mu$m, placing an upper limit of $1\times 10^{-5}$ on the systematic uncertainty of $g_{D_{5/2}}$.

Presence of trapping rf voltages may cause an ac-Zeeman effect that would perturb the resonance frequencies of the ground and excited states differently.  This effect may manifest itself either because the ion exhibits enhanced micromotion and oscillates at $\omega_{trap}$ along a magnetic field gradient or due to an ac current at $\omega_{trap}$ produced by the trapping electrodes, which would produce an oscillating magnetic field at the ion.  Either of these ac-Zeeman effects depend on the amplitude of the rf voltage that drives the trap.  Measuring $g_{D_{5/2}}$ at vastly different trap rf voltages, we find that an ac-Zeeman effect does not result in a statistically significant shift.  The measurements at low (high) trap rf voltage are plotted in blue (red) in Fig.~\ref{fig:gmeas}.  We can place a fractional error of $<10^{-6}$ on this effect.

Lastly, the experiment is triggered at a specific phase of the ac line voltage.   The small jitter in this line trigger would affect the measurement of $g_{D_{5/2}}$ below the $10^{-7}$ level.

A summary of these systematic effects can be found in Table~\ref{tab:syserror}.

\begin{table}[h]%The best place to locate the table environment is directly after its first reference in text
\caption{\label{tab:syserror}%
A summary of systematic error estimates for our experimental apparatuses. }
\begin{ruledtabular}
\begin{tabular}{lrr}
\textrm{Error Source}&
\textrm{$\Delta g_{D_{5/2}}$ ``Rod"}&
\textrm{$\Delta g_{D_{5/2}}$ ``Blade"}\\

\colrule
B Fluctuations &  $1.3\times 10^{-5}$ & $4.6\times 10^{-5}$ \\
B Gradient & $1\times 10^{-5}$ & $1\times 10^{-5}$\\
ac Zeeman Effect & $ 10^{-6}$ & $10^{-6}$\\
60Hz Jitter & $10^{-7}$ & $<10^{-7}$\\
\hline
Total & $1.6\times10^{-5}$ & $4.7\times10^{-5}$

\end{tabular}
\end{ruledtabular}
\end{table}

\section{Conclusions}
Using precision rf spectroscopy, we have measured the Land\'e  $g$ factor of the $5D_{5/2}$ level of $^{138}$Ba$^+$ to be 1.200~367$(6_{stat})(8_{sys})$ in the ``rod" apparatus and 1.200~388$(4_{stat})(21_{sys})$ in the ``blade" apparatus.  These independent measurements combine to yield a value of 1.200~371$(4_{stat})(6_{sys})$. The reduced $\chi^2$ of these 10 measurements was found to be 0.7.  This measurement offers a factor of 70 reduction in statistical uncertainty from the previous measurement \cite{kurz10} and is, to our knowledge, the most accurate measurement to date.   It should be noted that our group's previous result \cite{kurz10} disagrees with this new measurement by more than 3.5-$\sigma$.  The value in \cite{kurz10} is heavily shifted by a couple of outlying data points, which were included for completeness.  We reanalyzed this data, removing the outliers, and find a result of 1.2004(5), which is consistent with the improved measurement reported here.  It has also come to our attention that there is a new reported value of $g_{D_{5/2}}$ that differs in a statistically significant way from the result reported here \cite{lewty13}.  We feel that the simplicity of the methodology and analysis of the experiment reported here lends confidence to our result.

The improved measurement of $g_{D_{5/2}}$ should enable us to perform the measurement of the hyperfine intervals of $^{137}$Ba$^{+}$ once magnetic shielding is employed in the experimental setup, which is necessary for all of our future work.  This will reduce the most significant systematic error present in this experiment and will allow us to place better constraints in the future.  

\begin{acknowledgments}
The authors wish to thank John Wright, Richard Graham, Zichao Zhou, Chen-Kuan Chou, Nathan Kurz, Brent Graner, Jennie Chen, Eric Lindahl, and Blayne Heckel for helpful discussions.  Additional thanks to William Terrano and H. Erik Swanson for the use of a fluxgate magnetometer.  This research was supported by National Science Foundation grants 0906494 and 0904004.
\end{acknowledgments}


\begin{thebibliography}{16}
\expandafter\ifx\csname natexlab\endcsname\relax\def\natexlab#1{#1}\fi
\expandafter\ifx\csname bibnamefont\endcsname\relax
  \def\bibnamefont#1{#1}\fi
\expandafter\ifx\csname bibfnamefont\endcsname\relax
  \def\bibfnamefont#1{#1}\fi
\expandafter\ifx\csname citenamefont\endcsname\relax
  \def\citenamefont#1{#1}\fi
\expandafter\ifx\csname url\endcsname\relax
  \def\url#1{\texttt{#1}}\fi
\expandafter\ifx\csname urlprefix\endcsname\relax\def\urlprefix{URL }\fi
\providecommand{\bibinfo}[2]{#2}
\providecommand{\eprint}[2][]{\url{#2}}

\bibitem[{\citenamefont{Nagourney et~al.}(1986)\citenamefont{Nagourney,
  Sandberg, and Dehmelt}}]{dehmelt86}
\bibinfo{author}{\bibfnamefont{W.}~\bibnamefont{Nagourney}},
  \bibinfo{author}{\bibfnamefont{J.}~\bibnamefont{Sandberg}}, \bibnamefont{and}
  \bibinfo{author}{\bibfnamefont{H.}~\bibnamefont{Dehmelt}},
  \bibinfo{journal}{Phys. Rev. Lett.} \textbf{\bibinfo{volume}{56}},
  \bibinfo{pages}{2797} (\bibinfo{year}{1986}).

\bibitem[{\citenamefont{Gurell et~al.}(2007)\citenamefont{Gurell, Bi\'emont,
  Blagoev, Fivet, Lundin, Mannervik, Norlin, Quinet, Rostohar, Royen
  et~al.}}]{gurrell07}
\bibinfo{author}{\bibfnamefont{J.}~\bibnamefont{Gurell}},
  \bibinfo{author}{\bibfnamefont{E.}~\bibnamefont{Bi\'emont}},
  \bibinfo{author}{\bibfnamefont{K.}~\bibnamefont{Blagoev}},
  \bibinfo{author}{\bibfnamefont{V.}~\bibnamefont{Fivet}},
  \bibinfo{author}{\bibfnamefont{P.}~\bibnamefont{Lundin}},
  \bibinfo{author}{\bibfnamefont{S.}~\bibnamefont{Mannervik}},
  \bibinfo{author}{\bibfnamefont{L.-O.} \bibnamefont{Norlin}},
  \bibinfo{author}{\bibfnamefont{P.}~\bibnamefont{Quinet}},
  \bibinfo{author}{\bibfnamefont{D.}~\bibnamefont{Rostohar}},
  \bibinfo{author}{\bibfnamefont{P.}~\bibnamefont{Royen}},
  \bibnamefont{et~al.}, \bibinfo{journal}{Phys. Rev. A}
  \textbf{\bibinfo{volume}{75}}, \bibinfo{pages}{052506}
  (\bibinfo{year}{2007}),
  \urlprefix\url{http://link.aps.org/doi/10.1103/PhysRevA.75.052506}.

\bibitem[{\citenamefont{Lindroth and Ynnerman}(1993)}]{lindroth93}
\bibinfo{author}{\bibfnamefont{E.}~\bibnamefont{Lindroth}} \bibnamefont{and}
  \bibinfo{author}{\bibfnamefont{A.}~\bibnamefont{Ynnerman}},
  \bibinfo{journal}{Phys. Rev. A} \textbf{\bibinfo{volume}{47}},
  \bibinfo{pages}{961} (\bibinfo{year}{1993}),
  \urlprefix\url{http://link.aps.org/doi/10.1103/PhysRevA.47.961}.

\bibitem[{\citenamefont{Sahoo}(2006)}]{sahoo06quad}
\bibinfo{author}{\bibfnamefont{B.~K.} \bibnamefont{Sahoo}},
  \bibinfo{journal}{Phys. Rev. A} \textbf{\bibinfo{volume}{74}},
  \bibinfo{pages}{020501} (\bibinfo{year}{2006}),
  \urlprefix\url{http://link.aps.org/doi/10.1103/PhysRevA.74.020501}.

\bibitem[{\citenamefont{Lewty et~al.}(2012)\citenamefont{Lewty, Chuah, Cazan,
  Sahoo, and Barrett}}]{lewty12}
\bibinfo{author}{\bibfnamefont{N.~C.} \bibnamefont{Lewty}},
  \bibinfo{author}{\bibfnamefont{B.~L.} \bibnamefont{Chuah}},
  \bibinfo{author}{\bibfnamefont{R.}~\bibnamefont{Cazan}},
  \bibinfo{author}{\bibfnamefont{B.~K.} \bibnamefont{Sahoo}}, \bibnamefont{and}
  \bibinfo{author}{\bibfnamefont{M.~D.} \bibnamefont{Barrett}},
  \bibinfo{journal}{Opt. Express} \textbf{\bibinfo{volume}{20}},
  \bibinfo{pages}{21379} (\bibinfo{year}{2012}),
  \urlprefix\url{http://www.opticsexpress.org/abstract.cfm?URI=oe-20-19-21379}.

\bibitem[{\citenamefont{Beloy et~al.}(2008)\citenamefont{Beloy, Derevianko,
  Dzuba, Howell, Blinov, and Fortson}}]{howell08}
\bibinfo{author}{\bibfnamefont{K.}~\bibnamefont{Beloy}},
  \bibinfo{author}{\bibfnamefont{A.}~\bibnamefont{Derevianko}},
  \bibinfo{author}{\bibfnamefont{V.~A.} \bibnamefont{Dzuba}},
  \bibinfo{author}{\bibfnamefont{G.~T.} \bibnamefont{Howell}},
  \bibinfo{author}{\bibfnamefont{B.~B.} \bibnamefont{Blinov}},
  \bibnamefont{and} \bibinfo{author}{\bibfnamefont{E.~N.}
  \bibnamefont{Fortson}}, \bibinfo{journal}{Phys. Rev. A}
  \textbf{\bibinfo{volume}{77}}, \bibinfo{pages}{052503}
  (\bibinfo{year}{2008}).

\bibitem[{\citenamefont{Moore}(1908)}]{moore1908}
\bibinfo{author}{\bibfnamefont{B.~E.} \bibnamefont{Moore}},
  \bibinfo{journal}{Annalen der Physik} \textbf{\bibinfo{volume}{330}},
  \bibinfo{pages}{309} (\bibinfo{year}{1908}), ISSN \bibinfo{issn}{1521-3889},
  \urlprefix\url{http://dx.doi.org/10.1002/andp.19083300208}.

\bibitem[{\citenamefont{Back}(1923)}]{back1923}
\bibinfo{author}{\bibfnamefont{E.}~\bibnamefont{Back}},
  \bibinfo{journal}{Annalen der Physik} \textbf{\bibinfo{volume}{375}},
  \bibinfo{pages}{333} (\bibinfo{year}{1923}), ISSN \bibinfo{issn}{1521-3889},
  \urlprefix\url{http://dx.doi.org/10.1002/andp.19233750503}.

\bibitem[{\citenamefont{Curry}(2004)}]{curry04}
\bibinfo{author}{\bibfnamefont{J.~J.} \bibnamefont{Curry}},
  \bibinfo{journal}{Journal of Physical and Chemical Reference Data}
  \textbf{\bibinfo{volume}{33}}, \bibinfo{pages}{725} (\bibinfo{year}{2004}),
  \urlprefix\url{http://link.aip.org/link/?JPR/33/725/1}.

\bibitem[{\citenamefont{Kurz et~al.}(2010)\citenamefont{Kurz, Dietrich, Shu,
  Noel, and Blinov}}]{kurz10}
\bibinfo{author}{\bibfnamefont{N.}~\bibnamefont{Kurz}},
  \bibinfo{author}{\bibfnamefont{M.~R.} \bibnamefont{Dietrich}},
  \bibinfo{author}{\bibfnamefont{G.}~\bibnamefont{Shu}},
  \bibinfo{author}{\bibfnamefont{T.}~\bibnamefont{Noel}}, \bibnamefont{and}
  \bibinfo{author}{\bibfnamefont{B.~B.} \bibnamefont{Blinov}},
  \bibinfo{journal}{Phys. Rev. A} \textbf{\bibinfo{volume}{82}},
  \bibinfo{pages}{030501} (\bibinfo{year}{2010}),
  \urlprefix\url{http://link.aps.org/doi/10.1103/PhysRevA.82.030501}.

\bibitem[{\citenamefont{Olmschenk et~al.}(2007)\citenamefont{Olmschenk, Younge,
  Moehring, Matsukevich, Maunz, and Monroe}}]{olmschenk07}
\bibinfo{author}{\bibfnamefont{S.}~\bibnamefont{Olmschenk}},
  \bibinfo{author}{\bibfnamefont{K.~C.} \bibnamefont{Younge}},
  \bibinfo{author}{\bibfnamefont{D.~L.} \bibnamefont{Moehring}},
  \bibinfo{author}{\bibfnamefont{D.~N.} \bibnamefont{Matsukevich}},
  \bibinfo{author}{\bibfnamefont{P.}~\bibnamefont{Maunz}}, \bibnamefont{and}
  \bibinfo{author}{\bibfnamefont{C.}~\bibnamefont{Monroe}},
  \bibinfo{journal}{Phys. Rev. A} \textbf{\bibinfo{volume}{76}},
  \bibinfo{pages}{052314} (\bibinfo{year}{2007}).

\bibitem[{\citenamefont{Gulde}(2003)}]{gulde03}
\bibinfo{author}{\bibfnamefont{S.}~\bibnamefont{Gulde}}, Ph.D. thesis,
  \bibinfo{school}{University of Innsbruck}, \bibinfo{address}{Innsbruck,
  Austria} (\bibinfo{year}{2003}).

\bibitem[{\citenamefont{Noel et~al.}(2012)\citenamefont{Noel, Dietrich, Kurz,
  Shu, Wright, and Blinov}}]{noel12}
\bibinfo{author}{\bibfnamefont{T.}~\bibnamefont{Noel}},
  \bibinfo{author}{\bibfnamefont{M.~R.} \bibnamefont{Dietrich}},
  \bibinfo{author}{\bibfnamefont{N.}~\bibnamefont{Kurz}},
  \bibinfo{author}{\bibfnamefont{G.}~\bibnamefont{Shu}},
  \bibinfo{author}{\bibfnamefont{J.}~\bibnamefont{Wright}}, \bibnamefont{and}
  \bibinfo{author}{\bibfnamefont{B.~B.} \bibnamefont{Blinov}},
  \bibinfo{journal}{Phys. Rev. A} \textbf{\bibinfo{volume}{85}},
  \bibinfo{pages}{023401} (\bibinfo{year}{2012}),
  \urlprefix\url{http://link.aps.org/doi/10.1103/PhysRevA.85.023401}.

\bibitem[{\citenamefont{Wunderlich et~al.}(2007)\citenamefont{Wunderlich,
  Hannemann, K\"orber, H\"affner, Roos, H\"ansel, Blatt, and
  Schmidt-Kaler}}]{wunderlich07}
\bibinfo{author}{\bibfnamefont{C.}~\bibnamefont{Wunderlich}},
  \bibinfo{author}{\bibfnamefont{T.}~\bibnamefont{Hannemann}},
  \bibinfo{author}{\bibfnamefont{T.}~\bibnamefont{K\"orber}},
  \bibinfo{author}{\bibfnamefont{H.}~\bibnamefont{H\"affner}},
  \bibinfo{author}{\bibfnamefont{C.}~\bibnamefont{Roos}},
  \bibinfo{author}{\bibfnamefont{W.}~\bibnamefont{H\"ansel}},
  \bibinfo{author}{\bibfnamefont{R.}~\bibnamefont{Blatt}}, \bibnamefont{and}
  \bibinfo{author}{\bibfnamefont{F.}~\bibnamefont{Schmidt-Kaler}},
  \bibinfo{journal}{Journal of Modern Optics} \textbf{\bibinfo{volume}{54}},
  \bibinfo{pages}{1541} (\bibinfo{year}{2007}),
  \urlprefix\url{http://www.tandfonline.com/doi/abs/10.1080/09500340600741082}.

\bibitem[{\citenamefont{Knab et~al.}(1993)\citenamefont{Knab, Kn\"oll,
  Scheerer, and Werth}}]{knab93}
\bibinfo{author}{\bibfnamefont{H.}~\bibnamefont{Knab}},
  \bibinfo{author}{\bibfnamefont{K.}~\bibnamefont{Kn\"oll}},
  \bibinfo{author}{\bibfnamefont{F.}~\bibnamefont{Scheerer}}, \bibnamefont{and}
  \bibinfo{author}{\bibfnamefont{G.}~\bibnamefont{Werth}},
  \bibinfo{journal}{Zeitschrift f\"ur Physik D Atoms, Molecules and Clusters}
  \textbf{\bibinfo{volume}{25}}, \bibinfo{pages}{205} (\bibinfo{year}{1993}),
  ISSN \bibinfo{issn}{0178-7683}.

\bibitem[{\citenamefont{Lewty et~al.}(2013)\citenamefont{Lewty, Chuah, Cazan,
  Sahoo, and Barrett}}]{lewty13}
\bibinfo{author}{\bibfnamefont{N.~C.} \bibnamefont{Lewty}},
  \bibinfo{author}{\bibfnamefont{B.~L.} \bibnamefont{Chuah}},
  \bibinfo{author}{\bibfnamefont{R.}~\bibnamefont{Cazan}},
  \bibinfo{author}{\bibfnamefont{B.~K.} \bibnamefont{Sahoo}}, \bibnamefont{and}
  \bibinfo{author}{\bibfnamefont{M.~D.} \bibnamefont{Barrett}}
  (\bibinfo{year}{2013}), \bibinfo{note}{preprint},
  \urlprefix\url{http://arxiv.org/abs/1305.4453}.

\end{thebibliography}
\end{document}